# A lattice model describing scale effects in nonlinear elasticity of nano-inhomogeneities

Pier Luca Palla,* Stefano Giordano,† and Luciano Colombo‡

*Department of Physics of the University of Cagliari and*
*Istituto Officina dei Materiali del CNR (CNR-IOM) Unitá SLACS*
*Cittadella Universitaria, I-09042 Monserrato (Ca), Italy*

We present a procedure to map the constitutive laws of elasticity (both in the linear and nonlinear regime) onto a discrete atomic lattice and we apply the resulting elastic lattice model to investigate the strain field within an embedded nano-inhomogeneity. We prove that its elastic behavior at the nanoscale is governed by relevant atomistic effects. In particular, we demonstrate that such effects on the linear and nonlinear elastic properties are described by the same scaling exponent, in a large range of elastic contrast between the matrix and the nano-inhomogeneity. This suggests that the linear and nonlinear elastic behaviors of the composite system belong to the same universality class (at least within the nanometer length scale here investigated).

## I. INTRODUCTION

In modern nano-materials science, heterogeneous structures (like, e.g., nano-composites) are widely investigated because of their peculiar properties of large technological impact. The length scale at which their structural complexity occurs typically ranges from the microscale to the nanoscale, so that major conceptual difficulties arise by using continuum theories to investigate their physical behavior. In particular, this is the case of continuum elasticity theory that can hardly work at the nanoscale (where matter shows its atomistic structure) and, therefore, it is often inadequate in predicting many relevant mechanical properties of nano-structures. Another crucial aspect for understanding the elastic behavior of nano-systems is the nonlinearity of their constitutive equations: even this feature is often overlooked by standard applications of elasticity theory, since it leads to severe complications both from the theoretical and the numerical point of view. However, under many respects, the interplay between scale effects and nonlinear elastic response can be identified as a key issue in the mechanics of nano-composites.

The starting model system typically developed by continuum mechanics to describe heterogeneous structures is based on a single inhomogeneity, embedded in a given homogeneous matrix: this simple paradigmatic model is called "Eshelby configuration".[1,2] Continuum mechanics in general provides results about the elastic fields within and nearby an inhomogeneity which are inherently scale-invariant, i.e. they do not contain any explicit dependence upon the size of the inhomogeneity. Nevertheless, some models have been introduced to describe scale effects at surfaces and interfaces, which behave differently from their bulk counterparts.[3–6] They are based on the so-called Interface Stress Model (ISM) described by the displacement continuity condition and by the Young-Laplace equation for the stress behavior.[7,8] Such models have been applied to the Eshelby formalism with interface effects,[9,10] to stratified particles,[11] and to alloyed quantum dots.[12] Other models introducing scale effects are based upon nonlocal continuum field theories, which consider terms containing gradients of strain and rotation in the elastic energy density.[13]

In any case, we remark that the Eshelby configuration is affordable by continuum methods for just few selected combinations of elastically nonlinear matrix or inhomogeneity. The aim of this paper is to attempt a more general solution of this problem based on an elastic lattice model, addressing both scale effects and nonlinear elastic phenomena. In our approach we follow a different path where no educated guess on the actual constitutive behavior for the interface (or nonlocal continuum model) is assumed. Rather, we directly deduce the scale effects from atomic-scale features. It is important to remark that in this way we can deal with nonlinear properties and their scale effects.

In order to develop our formalism, we take into consideration a benchmark situation of paradigmatic importance, namely: a single circular inhomogeneity embedded in a given matrix, under remote loading and plain strain border conditions. The continuum Eshelby formalism provides the complete solution of this problem just in two cases: when both the inhomogeneity and the matrix are linear elastic,[14,15] and when a nonlinear inhomogeneity is embedded into a linear matrix.[16,17] These two situations are used to set up a nonlinear elastic lattice model, where the radius of the inhomogeneity could be varied. The goal is to provide a characterization of the scale effects on both the linear and the nonlinear features of the embedded particle. Moreover, we investigate two more configurations that are overwhelmingly difficult to investigate by the classical Eshelby theory, namely: linear inhomogeneity in nonlinear matrix and nonlinear inhomogeneity in nonlinear matrix.

The present elastic lattice model is developed in three conceptual steps: (i) we map continuum elasticity onto a discrete lattice; (ii) we introduce a suitable interatomic distance by means of which the notion of length scale is naturally introduced and, therefore, the possible onset of scale effects can be described; (iii) we translate an arbitrary continuum constitutive law (either linear or not) into a simple atomistic interaction potential which is eventually put at work on the above lattice. We apply the corresponding formalism to two-dimensional elastic problems, which could correspond (under plane strain or plane stress border conditions) to systems of actual current interest, like e.g.: dispersions of inhomogeneities or buried quantum dots,[18–22] graded structures,[23,24] complex interfaces,[25–28] nano-alloys or -composites[29,30]. For ease of implementation, we make use of simple two-body interaction potentials (harmonic, linearized and anharmonic) to govern the mechanics of the two-dimensional lattice. Our choice is motivated by the limited purpose of this work, which is basically a proof-of-concept investigation addressed to combine nonlinear effects with length-scale ones. However, we remark that, although simple, these potentials are just enough to describe several elastically different materials, since we can vary independently their elastic moduli. The extension to many-body potentials will be worked out next.

The structure of the paper is the following. In Section II we describe the continuum strain energy function and the corresponding interatomic interactions for the triangular lattice model. In Section III we present a brief outline of the two-dimensional Eshelby theory for the case of a linear or a nonlinear inhomogeneity embedded in a linear matrix. Finally, in Section IV, we describe the results obtained for the four Eshelby configurations above discussed.

## II. MAPPING CONTINUUM ELASTICITY ONTO A LATTICE

We develop our formalism for a two-dimensional triangular lattice of atoms, belonging to the hexagonal crystal symmetry. Our choice is motivated by the fact that such a structure is the only lattice exhibiting an isotropic linear elastic behavior, as indeed requested by the Eshelby theory. However, the conceptual device here introduced can be straightforwardly extended to three-dimensional lattices and/or arbitrary crystal symmetries.

The starting point is represented by the Neumann principle, namely:[31] the symmetry elements of any macroscopic tensor property of the crystal must include the symmetry elements of its point group. Accordingly, since the strain energy function (or elastic energy density) $\mathcal{U}(\hat{\varepsilon})$ for the hexagonal lattice is invariant under a rotation of $\pi/3$ about the principal axis (normal to the lattice plane), we can state that such a lattice is isotropically elastic in the linear regime and, therefore, its behavior is described by the Lamé coefficients $\lambda$ and $\mu$. On the other hand, when nonlinear elastic effects are taken into account, the resulting behavior is anisotropic, characterized by three independent elastic moduli $\Lambda_1$, $\Lambda_2$ and $\Lambda_3$.[32] Under these conditions, it can be easily proved that

$$\begin{aligned}\mathcal{U}(\hat{\varepsilon}) &= \frac{\lambda}{2}[\mathrm{Tr}(\hat{\varepsilon})]^2 + \mu\,\mathrm{Tr}(\hat{\varepsilon}^2)\\ &+ \Lambda_1(\varepsilon_{11}-\varepsilon_{22})[(\varepsilon_{11}-\varepsilon_{22})^2 - 12\varepsilon_{12}^2]\\ &+ \frac{1}{2}\Lambda_2\mathrm{Tr}(\hat{\varepsilon})[2\,\mathrm{Tr}(\hat{\varepsilon}^2) - \mathrm{Tr}(\hat{\varepsilon})^2]\\ &+ \frac{1}{2}\Lambda_3\mathrm{Tr}(\hat{\varepsilon})^3 \end{aligned} \tag{1}$$

where $\lambda = C_{12}$, $2\mu = C_{11} - C_{12}$, and

$$\Lambda_1 = \frac{1}{12}(C_{111} - C_{222}), \qquad \Lambda_2 = \frac{1}{4}(C_{222} - C_{112}),$$
$$\Lambda_3 = \frac{1}{12}(2C_{111} - C_{222} + 3C_{112}). \tag{2}$$

In Eq.(2) $C_{\alpha\beta}$ and $C_{\alpha\beta\gamma}$ are, respectively, the linear and nonlinear elastic constants as customarily defined in crystal elasticity.[32] We remark that the linear isotropy condition $C_{44} = (C_{11} - C_{12})/2$ is always satisfied. The infinitesimal strain tensor $\hat{\varepsilon} = \frac{1}{2}(\vec{\nabla}\vec{u} + \vec{\nabla}\vec{u}^{\mathrm{T}})$ is represented by a symmetric matrix with elements $\varepsilon_{11} = \frac{\partial u_1}{\partial x_1}$, $\varepsilon_{22} = \frac{\partial u_2}{\partial x_2}$ and $\varepsilon_{12} = \frac{1}{2}\left(\frac{\partial u_1}{\partial x_2} + \frac{\partial u_2}{\partial x_1}\right)$ where the functions $u_1(x_1, x_2)$ and $u_2(x_1, x_2)$ correspond to the planar displacement $\vec{u} = (u_1, u_2)$. In this work we adopt everywhere the small strain formalism in order to allow the comparison with the Eshelby results: in fact, the standard Eshelby configurations can not be solved within the finite elasticity theory.

A fully isotropic system described by a strain energy function

$$\mathcal{U}(\hat{\varepsilon}) = \frac{\lambda}{2}\mathrm{Tr}(\hat{\varepsilon})^2 + \mu\,\mathrm{Tr}(\hat{\varepsilon}^2) + \mathsf{e}\,\mathrm{Tr}(\hat{\varepsilon})\mathrm{Tr}(\hat{\varepsilon}^2) + \mathsf{f}\,\mathrm{Tr}(\hat{\varepsilon})^3 \tag{3}$$

is obtained under the condition $C_{111} = C_{222}$. In Eq.(3) the nonlinear behavior is described by the two coefficients $\mathsf{e}$ and $\mathsf{f}$ which are easily calculated as

$$\mathsf{e} = \frac{1}{4}(C_{111} - C_{112}) \qquad \mathsf{f} = \frac{1}{4}(C_{112} - \frac{1}{3}C_{111}) \tag{4}$$

If we now admit that each site of the triangular lattice is occupied by an atom, then we can express $\mathcal{U}(\hat{\varepsilon})$ in terms of a suitable interatomic potential. We assume that such a potential depends upon the distance vectors $\vec{r}_{ij}$ between each $ij$-th pair of atoms and we indicate $\vec{r}_{ij}^{\,0}$ such distances in the reference (i.e. unstrained) configuration. We further assume that the interatomic interactions are described by a simple harmonic spring between two next neighbor atoms, respectively placed at $\vec{r}_i$ and $\vec{r}_j$. The corresponding pairwise *harmonic* potential energy is

$$U_h = \frac{1}{2}\kappa_h(r_{ij} - r_0)^2 \tag{5}$$

where $r_{ij} = \mid \vec{r}_i - \vec{r}_j \mid$, $\kappa_h$ is the spring constant, and $r_0$ is the first next neighbor equilibrium distance. $U_h$ is intended to mimic bond-stretching interactions.

If the system is subjected to a displacement field $\vec{u}$ the deformed position of the $i$-th atom is written as $\vec{r}_i = \vec{r}_i^{\,0} + \vec{u}\left(\vec{r}_i^{\,0}\right)$; therefore, by considering $\vec{r}_{ij} = \vec{r}_{ij}^{\,0} + \Delta\vec{u}_{ij}$ with $\Delta\vec{u}_{ij} = \vec{u}(\vec{r}_i^{\,0}) - \vec{u}(\vec{r}_j^{\,0})$, the potential energy for the $ij$-th pair can be expanded in powers of $\vec{u}$

$$U_h = \frac{1}{2}\kappa_h \left(\vec{n}_{ij} \cdot \Delta\vec{u}_{ij}\right)^2 + \mathcal{O}(u^3) \tag{6}$$

where $\vec{n}_{ij} = \vec{r}_{ij}^{\,0}/r_0$ is the direction of the $i - j$ bond. It is easily verified that the linear elastic moduli $C_{\alpha\beta}$ are proportional to the potential parameter $\kappa_h$. Moreover, through the $\mathcal{O}(u^3)$ term in Eq.(6), the harmonic interaction affects also the nonlinear behavior: the $C_{\alpha\beta\gamma}$ moduli will be proportional to $\kappa_h$ as well.

If we are interested in a purely linear elastic system, then we can truncate Eq.(6) to just the second-order term, so obtaining a simple linear interaction potential. This suggests to consider another different contribution to the potential energy, hereafter named *linearized* spring and assuming the form

$$U_l \;=\; \mathcal{L}\left[\frac{1}{2}\kappa_l \left(r_{ij} - r_0\right)^2\right] = \frac{1}{2}\kappa_l \left(\vec{n}_{ij} \cdot \Delta\vec{u}_{ij}\right)^2 \tag{7}$$

In Eq.(7) we have introduced a new spring constant $k_l$ ($l$ means "linearized") and the linearization operator $\mathcal{L}$. In order to further generalize our scheme, we introduce an *anharmonic* term as well

$$U_a \;=\; \frac{1}{3}\frac{\kappa_a}{r_0}\left(r_{ij} - r_0\right)^3 \tag{8}$$

The resulting potential energy of the lattice is therefore written as

$$U \;=\; U_0 + \frac{1}{2}\sum_{ij}\left[U_l(r_{ij}) + U_h(r_{ij}) + U_a(r_{ij})\right] \tag{9}$$

where $U_0$ is a constant. It is important to remark that the *linearized* terms affect only the linear elastic moduli $C_{\alpha\beta}$; the *harmonic* terms affect both the linear $C_{\alpha\beta}$ and the nonlinear $C_{\alpha\beta\gamma}$ elastic constants; and, finally, the *anharmonic* terms affect only the nonlinear moduli $C_{\alpha\beta\gamma}$.

In summary, we have proved that a triangular lattice described by the atomistic potential energy given in Eq.(9) can be treated, from the continuum point of view, by the strain energy function given in Eq.(1), where the linear and nonlinear elastic moduli are provided by the following synopsis

$$\begin{aligned}
C_{11} &= \frac{3\sqrt{3}}{4}(\kappa_l + \kappa_h) \\
C_{12} &= \frac{\sqrt{3}}{4}(\kappa_l + \kappa_h) \\
C_{111} &= \frac{9\sqrt{3}}{16}\kappa_h + \frac{9\sqrt{3}}{8}\kappa_a \\
C_{222} &= \frac{3\sqrt{3}}{16}\kappa_h + \frac{11\sqrt{3}}{8}\kappa_a \\
C_{112} &= -\frac{5\sqrt{3}}{16}\kappa_h + \frac{3\sqrt{3}}{8}\kappa_a
\end{aligned} \tag{10}$$

Sometimes it is useful to write the previous results in terms of the Lagrangian strain $\hat{\eta} = \frac{1}{2}(\vec{\nabla}\vec{u} + \vec{\nabla}\vec{u}^{\mathrm{T}} + \vec{\nabla}\vec{u}^{\mathrm{T}}\vec{\nabla}\vec{u})$. While $\hat{\varepsilon}$ takes into account only possible physical nonlinearity features (i.e. a nonlinear stress-strain dependence observed in regime of small deformation), $\hat{\eta}$ describes any possible source of nonlinearity, including both physical and geometrical (large deformation) ones.[2] By using the Lagrangian strain $\hat{\eta}$, the strain energy function is given by the very same Eqs.(1) and (2), where $\hat{\varepsilon}$ is replaced by $\hat{\eta}$ and $C_{\alpha\beta\gamma}$ by the Lagrangian third-order moduli $C^{\mathcal{L}}_{\alpha\beta\gamma}$. By imposing the identity $U(\hat{\varepsilon}) = U(\hat{\eta})$ (where the Lagrangian strain can be written in term of the small strain by $\hat{\eta} = \hat{\varepsilon} + \frac{1}{2}\hat{\epsilon}^2$) we obtain the conversion rules: $C^{\mathcal{L}}_{111} = C_{111} - 3(2\mu + \lambda)$, $C^{\mathcal{L}}_{222} = C_{222} - 3(2\mu + \lambda)$, $C^{\mathcal{L}}_{112} = C_{112} - \lambda$ (the linear moduli $\lambda$, $\mu$, $C_{11}$ and $C_{12}$ remain unperturbed). Anyway, throughout all the paper we will use small deformations and the nonlinear moduli $C_{\alpha\beta\gamma}$.

## III. OUTLINE OF ESHELBY THEORY

The Eshelby theory provides a fundamental result, namely: the strain field within both a linear[14,15] or a nonlinear[16,17] inhomogeneity shown in Fig.1 is uniform (when the matrix is linear). Typically, it is assumed that no bonding failures occur at the interface when the structure is placed in an equilibrated state of infinitesimal elastic strain. So, the boundary conditions require that both the vector displacement and the normal stress be continuous across the interface.

If we identify the linear Lamé coefficients of the inhomogeneity by $\mu^{(2)}$ and $\lambda^{(2)}$, and its nonlinear constants as $\mathsf{e}$ and $\mathsf{f}$, then Eq.(3) supplies, by derivation, the following stress-strain relation

$$\begin{aligned}
\hat{T}^{(2)}(\hat{\varepsilon}^{(2)}) \;=\;& 2\mu^{(2)}\hat{\varepsilon}^{(2)} + \lambda^{(2)}\mathrm{Tr}(\hat{\varepsilon}^{(2)})\hat{I} \\
&+ 2\mathsf{e}\mathrm{Tr}\left(\hat{\varepsilon}^{(2)}\right)\hat{\varepsilon}^{(2)} + \mathsf{e}\mathrm{Tr}\left[(\hat{\varepsilon}^{(2)})^2\right]\hat{I} \\
&+ 3\mathsf{f}\mathrm{Tr}^2\left(\hat{\varepsilon}^{(2)}\right)\hat{I}
\end{aligned} \tag{11}$$

On the other hand, according to the Eshelby theory we can write[14–17]

$$\hat{\varepsilon}^{(2)} - \hat{S}\hat{\varepsilon}^{(2)} + \hat{S}\left(\hat{C}^{(1)}\right)^{-1}\hat{T}^{(2)}(\hat{\varepsilon}^{(2)}) = \hat{\varepsilon}^{\infty} \tag{12}$$

where $\hat{C}^{(1)}$ is the matrix stiffness tensor (with moduli $\mu^{(1)}$ and $\lambda^{(1)}$); $\hat{S}$ is the Eshelby tensor for a circular inhomogeneity; $\hat{\varepsilon}^{(2)}$ and $\hat{\varepsilon}^{\infty}$ represent the strain within the inhomogeneity and the remotely applied strain, respectively.

By replacing Eq.(11) into Eq.(12) we obtain the implicit equation for the internal field $\hat{\varepsilon}^{(2)}$

$$\begin{aligned}
\hat{\varepsilon}^{\infty} \;=\;& A\hat{\varepsilon}^{(2)} + B\mathrm{Tr}(\hat{\varepsilon}^{(2)})\hat{I} + C\mathrm{Tr}(\hat{\varepsilon}^{(2)})\hat{\varepsilon}^{(2)} \\
&+ D\mathrm{Tr}\left[(\hat{\varepsilon}^{(2)})^2\right]\hat{I} + E\mathrm{Tr}^2\left(\hat{\varepsilon}^{(2)}\right)\hat{I}
\end{aligned} \tag{13}$$

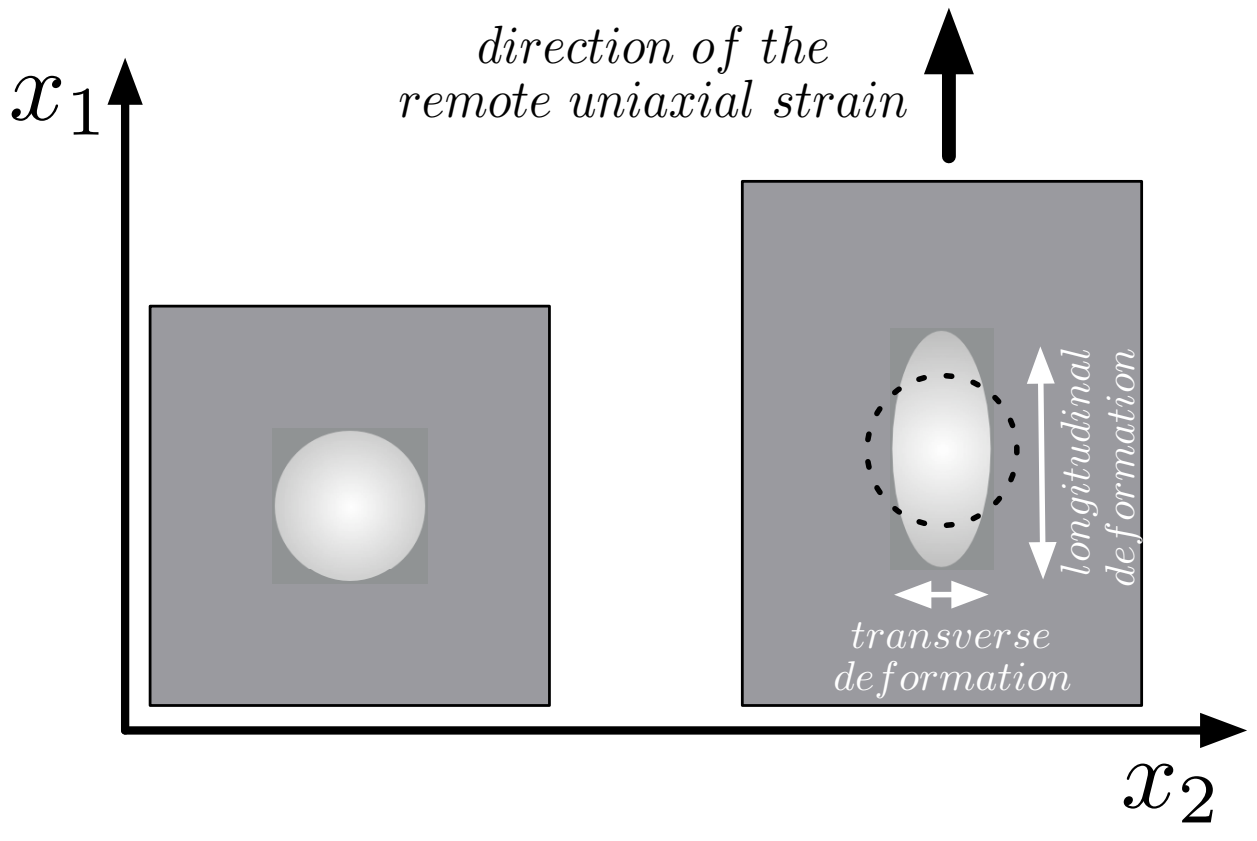

FIG. 1: (color online) Schematic representation of the matrix/inhomogeneity system in the unstrained (left) and strained (right) configuration. The matrix (inhomogeneity) is represented by the dark (light) gray shaded area. The strained configuration is obtained applying remotely an uniaxial strain along the $x_1$ direction. The resulting transverse and longitudinal deformation of the inhomogeneity is shown on the right with respect to its reference shape (dashed line). The corresponding strain tensor components are defined in the text.

where $\hat{I}$ is the identity operator and

$$\begin{aligned} A &= 1 - \frac{\lambda^{(1)} + 3\mu^{(1)}}{2(\lambda^{(1)} + 2\mu^{(1)})}\left(1 - \frac{\mu^{(2)}}{\mu^{(1)}}\right) \\ B &= \frac{2(\lambda^{(2)} - \lambda^{(1)}) + \left(1 - \frac{\mu^{(2)}}{\mu^{(1)}}\right)\left(\lambda^{(1)} + \mu^{(1)}\right)}{4(\lambda^{(1)} + 2\mu^{(1)})} \\ C &= \frac{1}{2\mu^{(1)}}\frac{\lambda^{(1)} + 3\mu^{(1)}}{\lambda^{(1)} + 2\mu^{(1)}}\mathsf{e} \\ D &= \frac{1}{2}\frac{\mathsf{e}}{\lambda^{(1)} + 2\mu^{(1)}} \\ E &= \frac{1}{2}\frac{3\mathsf{f}}{\lambda^{(1)} + 2\mu^{(1)}} - \frac{\lambda^{(1)} + \mu^{(1)}}{4\mu^{(1)}}\frac{\mathsf{e}}{\lambda^{(1)} + 2\mu^{(1)}} \end{aligned} \quad (14)$$

are constant parameters. In order to solve Eq.(13) for $\hat{\varepsilon}^{(2)}$ (up to the second order in $\hat{\varepsilon}^{\infty}$), it is useful to calculate the quantities $\hat{\varepsilon}^{\infty}$, $\mathrm{Tr}(\hat{\varepsilon}^{\infty})\hat{I}$, $\mathrm{Tr}(\hat{\varepsilon}^{\infty})\hat{\varepsilon}^{\infty}$, $\mathrm{Tr}\left[(\hat{\varepsilon}^{\infty})^2\right]\hat{I}$ and $\mathrm{Tr}^2\left(\hat{\varepsilon}^{\infty}\right)\hat{I}$ in terms of $\hat{\varepsilon}^{(2)}$. They can be arranged in matrix form as follows

$$\begin{pmatrix} \hat{\varepsilon}^{\infty} \\ \mathrm{Tr}(\hat{\varepsilon}^{\infty})\hat{I} \\ \mathrm{Tr}(\hat{\varepsilon}^{\infty})\hat{\varepsilon}^{\infty} \\ \mathrm{Tr}\left[(\hat{\varepsilon}^{\infty})^2\right]\hat{I} \\ \mathrm{Tr}^2\left(\hat{\varepsilon}^{\infty}\right)\hat{I} \end{pmatrix} = M \begin{pmatrix} \hat{\varepsilon}^{(2)} \\ \mathrm{Tr}(\hat{\varepsilon}^{(2)})\hat{I} \\ \mathrm{Tr}(\hat{\varepsilon}^{(2)})\hat{\varepsilon}^{(2)} \\ \mathrm{Tr}\left[(\hat{\varepsilon}^{(2)})^2\right]\hat{I} \\ \mathrm{Tr}^2\left(\hat{\varepsilon}^{(2)}\right)\hat{I} \end{pmatrix} \quad (15)$$

where

$$M = \begin{pmatrix} A & B & C & D & E \\ 0 & A+2B & 0 & 2D & C+2E \\ 0 & 0 & A(A+2B) & 0 & B(A+2B) \\ 0 & 0 & 0 & A^2 & 2B(A+B) \\ 0 & 0 & 0 & 0 & (A+2B)^2 \end{pmatrix} \quad (16)$$

By inverting the above system of equations, we eventually get the expression of the internal strain field $\hat{\varepsilon}^{(2)}$ as a function of the applied strain $\hat{\varepsilon}^{\infty}$

$$\begin{aligned} \hat{\varepsilon}^{(2)} &= \frac{\hat{\varepsilon}^{\infty}}{A} - \frac{B}{A(A+2B)}\mathrm{Tr}(\hat{\varepsilon}^{\infty})\hat{I} \\ &- \frac{1}{A^2(A+2B)}\left(C\mathrm{Tr}(\hat{\varepsilon}^{\infty})\hat{\varepsilon}^{\infty} + D\mathrm{Tr}\left[(\hat{\varepsilon}^{\infty})^2\right]\hat{I}\right) \\ &+ \frac{2B(A+B)(C+D) - EA^2}{A^2(A+2B)^3}\mathrm{Tr}^2\left(\hat{\varepsilon}^{\infty}\right)\hat{I} \end{aligned} \quad (17)$$

The applied homogeneous uniaxial elongation shown in Fig.1 is described by the strain tensor

$$\hat{\varepsilon}^{\infty} = \begin{pmatrix} \epsilon & 0 \\ 0 & 0 \end{pmatrix} \quad (18)$$

where $\epsilon$ is a scalar parameter describing the intensity of the uniaxial deformation. Eq.(17) assumes the form

$$\hat{\varepsilon}^{(2)} = \begin{pmatrix} \varepsilon_l & 0 \\ 0 & \varepsilon_t \end{pmatrix} = \begin{pmatrix} L^I\epsilon + L^{II}\epsilon^2 & 0 \\ 0 & T^I\epsilon + T^{II}\epsilon^2 \end{pmatrix} \quad (19)$$

where we have introduced the simplified notation $\varepsilon_{11} = \varepsilon_l = L^I\epsilon + L^{II}\epsilon^2$ and $\varepsilon_{22} = \varepsilon_t = T^I\epsilon + T^{II}\epsilon^2$ to indicate the fractional elongations along the longitudinal and the transverse directions, respectively (see Fig.1). Both $\varepsilon_l$ and $\varepsilon_t$ are quadratic functions of the remotely applied strain $\epsilon$ and the four corresponding coefficients ($L^I$ and $T^I$ for the linear response and $L^{II}$ and $T^{II}$ for the non-linear one) are the key quantities of this elastic problem. They are straightforwardly calculated as

$$L^I = \frac{A+B}{A(A+2B)} \quad (20)$$

$$T^I = \frac{-B}{A(A+2B)} \quad (21)$$

$$L^{II} = \frac{-(C+D)(A^2+2AB+2B^2) - EA^2}{A^2(A+2B)^3} \quad (22)$$

$$T^{II} = -\frac{A^2(D+E) + 2B(A+B)(D-C)}{A^2(A+2B)^3} \quad (23)$$

If the inhomogeneity is linear (i.e. $\mathsf{e} = 0$ and $\mathsf{f} = 0$), then we get $L^{II} = 0$ and $T^{II} = 0$ and the original linear Eshelby result is recovered.[14,15]

## IV. THE ELASTIC LATTICE MODEL AT WORK

In order to solve the Eshelby problem atomistically, we have selected two different elastic media: a fully isotropic linear material with $C_{111} = C_{222} = C_{112} = 0$ and an isotropic nonlinear one with $C_{111} = C_{222}$. The linear material is described by the set of parameters $\kappa_l = K$, $\kappa_h = 0$ and $\kappa_a = 0$, where $K$ is a constant governing the elastic stiffness. The resulting elastic behavior is thus given by the following moduli

$$
\begin{aligned}
C^l_{11} &= \frac{3\sqrt{3}}{4}K \\
C^l_{12} &= \frac{\sqrt{3}}{4}K \\
C^l_{111} &= 0 \\
C^l_{222} &= 0 \\
C^l_{112} &= 0
\end{aligned} \tag{24}
$$

The nonlinear material is described by setting $\kappa_l = 0$, $\kappa_h = K$ and $\kappa_a = \frac{3}{2}K$. Interactions are therefore composed by an *harmonic* term (affecting both the linear as the nonlinear elastic behavior) and by an *anharmonic* term (affecting only the nonlinear features), tailored to obtain an isotropic behavior. The resulting elastic moduli are

$$
\begin{aligned}
C^{nl}_{11} &= \frac{3\sqrt{3}}{4}K \\
C^{nl}_{12} &= \frac{\sqrt{3}}{4}K \\
C^{nl}_{111} &= \frac{9}{4}\sqrt{3}K \\
C^{nl}_{222} &= \frac{9}{4}\sqrt{3}K \\
C^{nl}_{112} &= \frac{\sqrt{3}}{4}K
\end{aligned} \tag{25}
$$

The isotropy is confirmed by the validity of the relation $C^{nl}_{111} = C^{nl}_{222}$.

In order to avoid the formation of a disordered interface at the matrix/inhomogeneity boundary and the resulting effects on the elastic behavior of the system,[33,34] we have chosen the same equilibrium distance $r_0$ and the same crystallographic orientation for both the inhomogeneity and matrix materials. Moreover, we have set $r_0 = 3.4$ Å, which is a reasonable interatomic distance for ceramic or covalently bonded compounds. In these conditions we have no lattice mismatch at the interface. This is crucially important in order to isolate the length scale effects from those possibly introduced by the disorder near the interface. Furthermore, the interaction between atoms belonging to the different phases (matrix and inhomogeneity) has been described by just a *linearized* spring with a constitutive parameter obtained through the geometric mean of the stiffness of the two adjacent materials (Lorentz-Berthelot rule[35]): this is a customarily used rule in multi-phases molecular dynamic studies. Moreover, we have proved that the present results do not really depend on this choice (provided that such a constitutive parameter varies in the range given by the values corresponding to the two phases).

We have applied to the system described above a set of uniaxial elongations in the longitudinal direction (see Fig.1) with $-0.01 \leq \epsilon \leq 0.01$. For each value of the deformation we have calculated the internal longitudinal and transverse strain, as defined in Eq.(19). In particular, for each deformed sample we have computed the displacement field inside the inhomogeneity along the direction parallel to the load (i.e. the longitudinal displacement $u_1(x_1, x_2)$) and the displacement field in the direction perpendicular to the load (i.e. the transverse displacement $u_2(x_1, x_2)$). We have so computed the longitudinal strain through the relation $\varepsilon_{11} = \varepsilon_l = \frac{\partial u_1}{\partial x_1}$ and the transverse strain through $\varepsilon_{22} = \varepsilon_t = \frac{\partial u_2}{\partial x_2}$. This analysis has been performed for different values of the *elastic contrast* between the matrix and the inhomogeneity, defined as $\log_2(K_{mat}/K_{inc})$, where $K_{mat}$ and $K_{inc}$ are the elastic stiffness parameters of the matrix and inhomogeneity, respectively, entering in Eqs.(24) and (25). A positive (negative) contrast means that the matrix is stiffer (softer) than the inhomogeneity. Moreover, all the simulations are been repeated for several values of the radius $R$ of the inhomogeneity in order to study the scale effects.

In the present simulations we have described the embedded nano-inhomogeneity by a simulation cell containing as many as 144000 atoms, which corresponds to a square box of length 120 nm (i.e. a cell size 20 larger than the radius $R$ of the inhomogeneity). Asymptotic Boundary Conditions (ABCs) have been adopted, namely: we have calculated by the Eshelby theory the displacement field due to the applied remote deformation ($\hat{\varepsilon}^{\infty}$) of an infinite two-dimensional elastic matrix containing a central single inhomogeneity; then, we have arranged all the atoms of the square simulation box according to the predicted displacement field. In this configuration, the atoms near the cell boundaries are subjected to a system of forces (maintaining the present configuration) calculated by the atomistic interaction model. After the application of the ABCs, the system has been relaxed through dumped dynamics in order to allow for the relaxation of the internal degrees of freedom. The convergence criterion has been set so as to have the final interatomic forces not larger than $10^{-10}$ eV/Å.

### A. Linear inhomogeneity in a linear matrix

This case was modeled by Eq.(24) for both the matrix and the inhomogeneity. We have calculated the longitudinal and transverse components of the internal strain as function of the elastic contrast for an inhomogeneity with radius as small as $R = 10$ Å (corresponding to 30 atoms). It is proved that the atomistic model provides a uniform internal strain field as predicted by continuum: a qualitative important feature standing for the reliability of the present elastic lattice model.

As shown by Eq.(19), the strain components are defined by the $L^I$ and $T^I$ coefficients: results from the

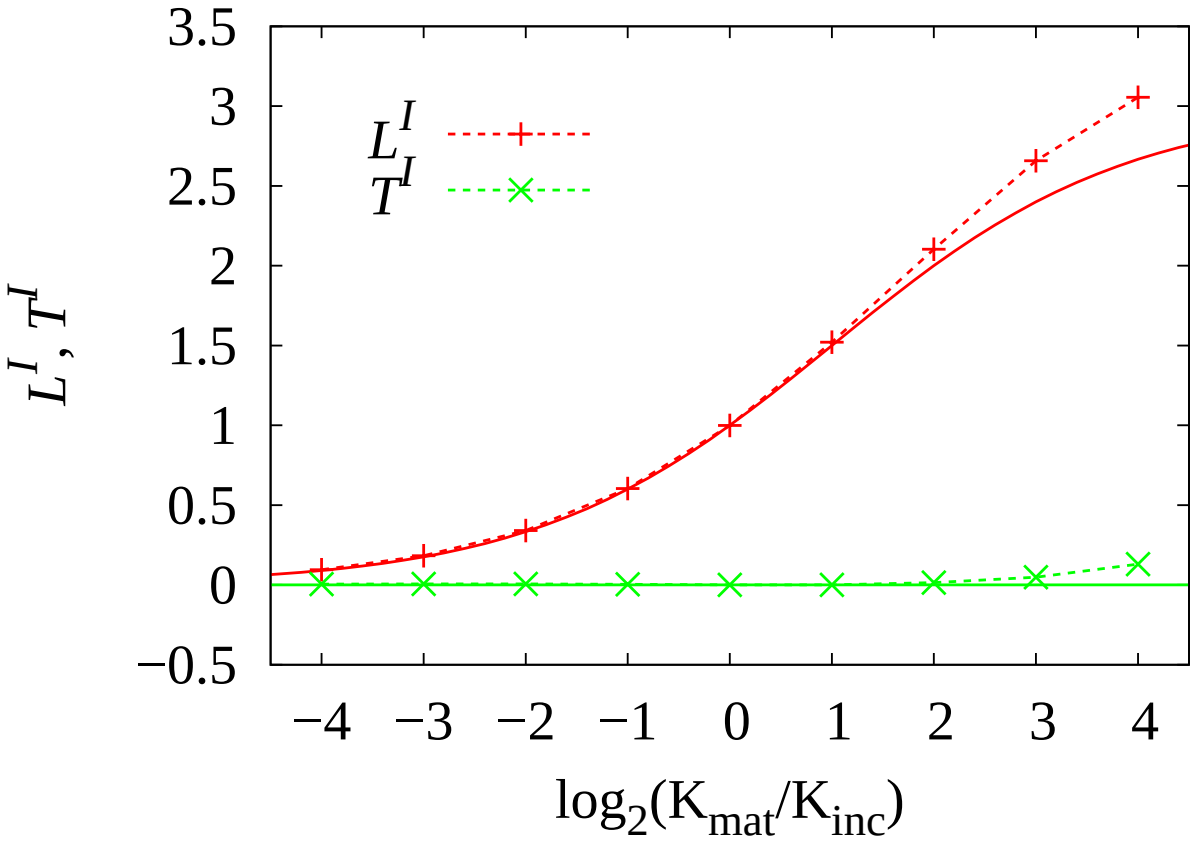

FIG. 2: (color online) Elastic behavior of a linear inhomogeneity embedded in a linear matrix under remote load as summarized by the longitudinal ($L^I$) and transverse ($T^I$) coefficients, respectively given in Eqs.(20) and (21). The solid lines represent the results of the continuum Eshelby theory. Dashed lines with + (×) symbols represent the atomic result for the longitudinal (transverse) coefficient.

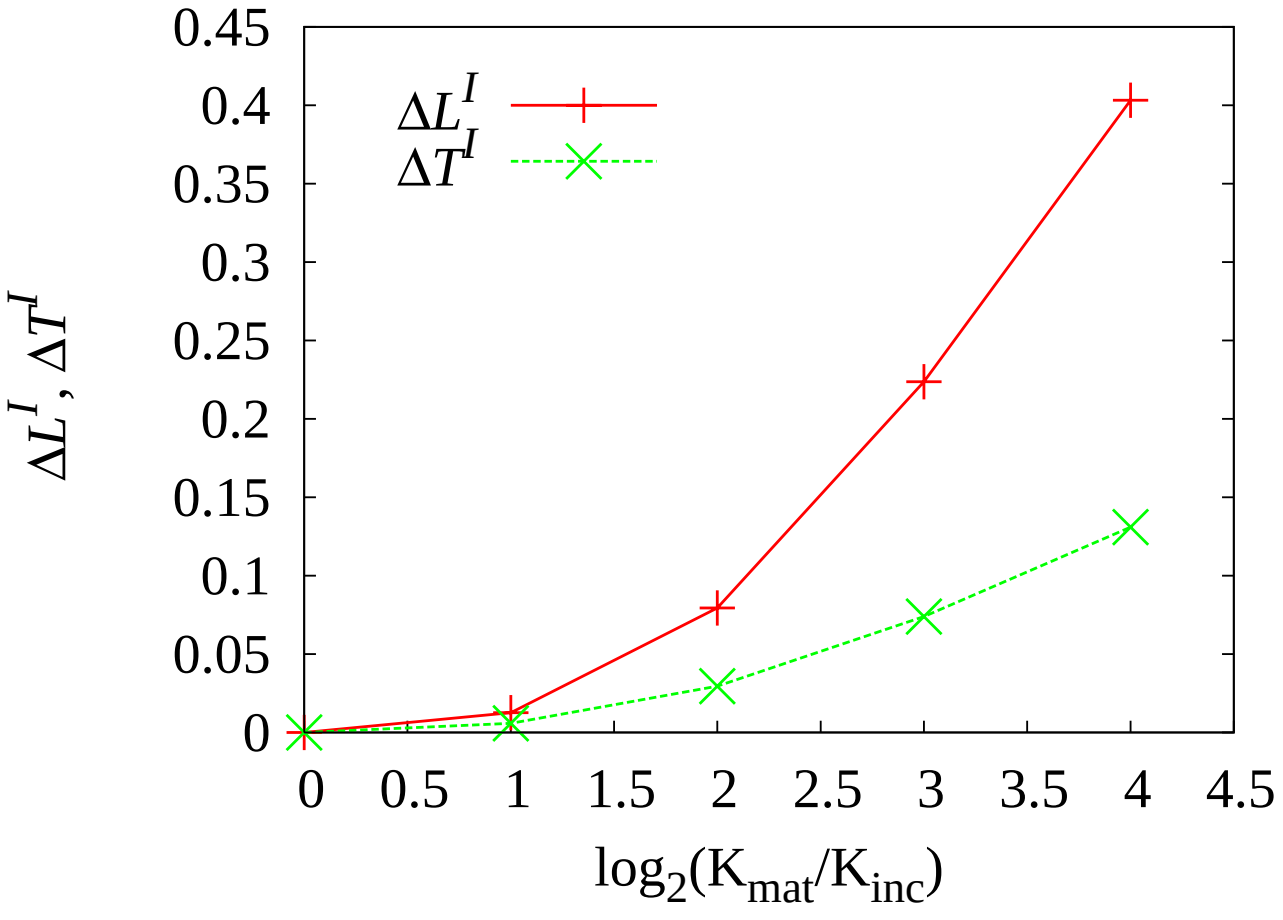

FIG. 3: (color online) Differences $\Delta L^I$ (full line with + symbols) and $\Delta T^I$ (dashed line with × symbols) between atomistic and continuum results for the longitudinal and transverse linear coefficients, respectively, versus the elastic contrast.

continuum Eshelby theory and the atomistic model are compared in Fig.2. The zero contrast data correspond to a homogeneous material. Therefore, as expected in this case, the external strain field is equal to the internal one or, equivalently: $L^I = 1$ and $T^I = 0$ for the loading condition shown in Fig.1. Interesting enough, when the contrast is positive (i.e. when the inhomogeneity is softer than the hosting matrix) atomistic data slightly differ from the continuum prediction, while for negative contrast a perfect agreement between the two approaches

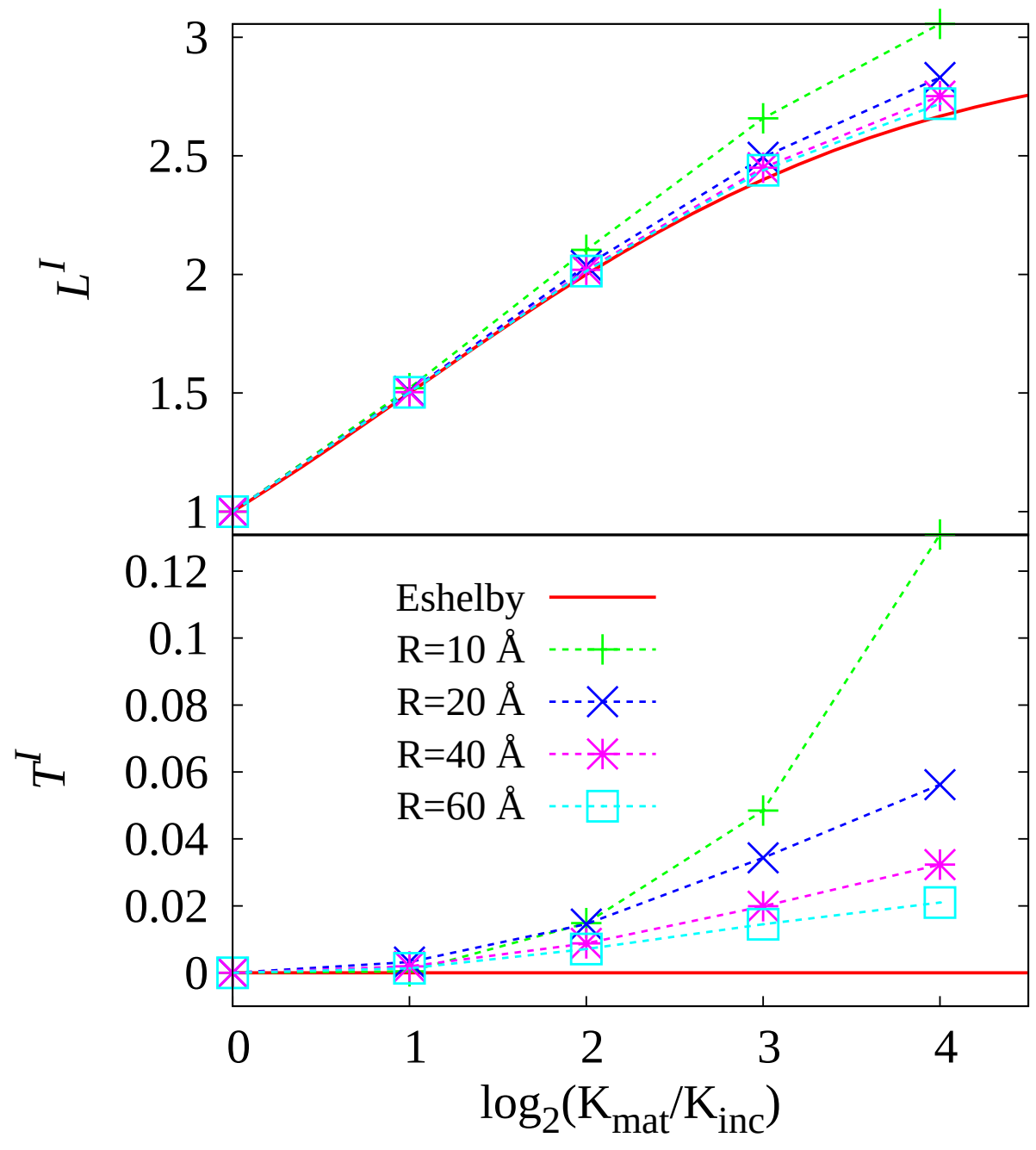

FIG. 4: (color online) Longitudinal (top panel) and transverse (bottom panel) linear coefficient as a function of the elastic contrast, calculated for increasing values of the inhomogeneity radius $R$. The solid lines represent the Eshelby predictions, while dashed lines with symbols represent the atomistic results.

is observed. This effect increases with the contrast as shown in Fig.3.

The disagreement between the continuum theory predictions and the atomistic results has been further investigated by varying the radius of the inhomogeneity. We have found that it vanishes by increasing the radius of the inhomogeneity, as shown in Fig.4. Therefore, we attribute this effect to atomic-scale features, not properly taken into account by the Eshelby theory. This suggests that by the present atomistic simulations we have set a lower limit of validity for the Eshelby theory, as far as the length scale is concerned. From Figs.3 and 4 it is found that such scale effects are much stronger for the longitudinal coefficient $L^I$ than for the transverse one $T^I$.

### B. Nonlinear inhomogeneity in a linear matrix

In this case we have modeled the inhomogeneity by the nonlinear isotropic elastic model represented by Eq.(25) setting $K = K_{inc}$. The matrix is described, in turn, by a linear material with $K = K_{mat}$ as before (see Eq.(24)). As before the inhomogeneity radius is $R = 10$ Å and the resulting internal strain field was found to be uniform.

In Fig.5 we report the longitudinal and transverse coefficients of Eq.(19) for several values of the elastic con-trast. By comparing Fig.2 to Fig.5, we note that the coefficients $L^I$ and $T^I$ are just the same. This means that any linear feature of the coupled inhomogeneity/matrix system is not affected by a possibly nonlinear inhomogeneity. As for the nonlinear $L^{II}$ and $T^{II}$ coefficient, we found once again a perfect agreement between atomistic and Eshelby results under the condition that the inhomogeneity is stiffer than the matrix (negative values of the contrast). On the other hand, atomistic effects are present in the case of positive contrast, as shown in Fig.6.

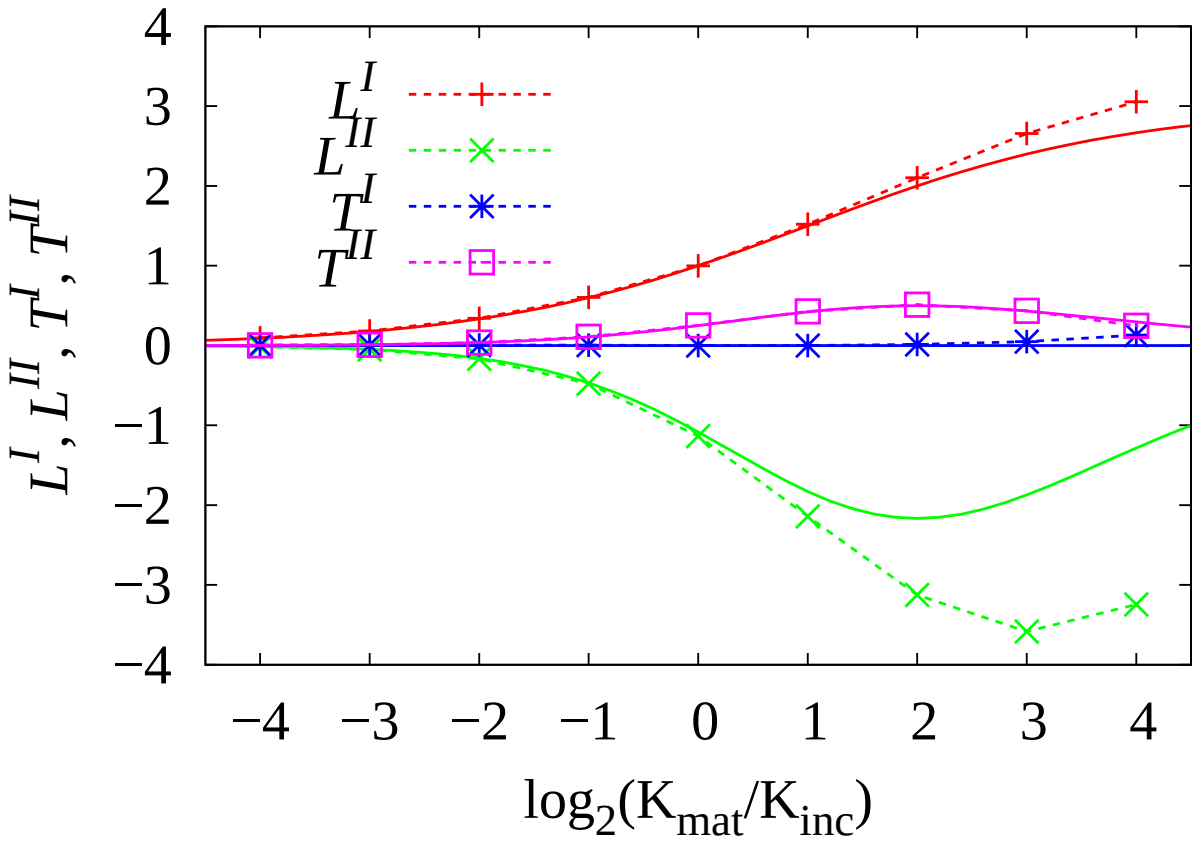

FIG. 5: (color online) Elastic behavior of a nonlinear inhomogeneity embedded in a linear matrix under remote load as summarized by the longitudinal ($L^I$ and $L^{II}$) and transverse ($T^I$ and $T^{II}$) coefficients, given in Eqs.(20-23). The solid lines represent the results of the continuum Eshelby theory. Dashed lines with + and × (□ and ∗) symbols represent the atomic result for the longitudinal (transverse) coefficients.

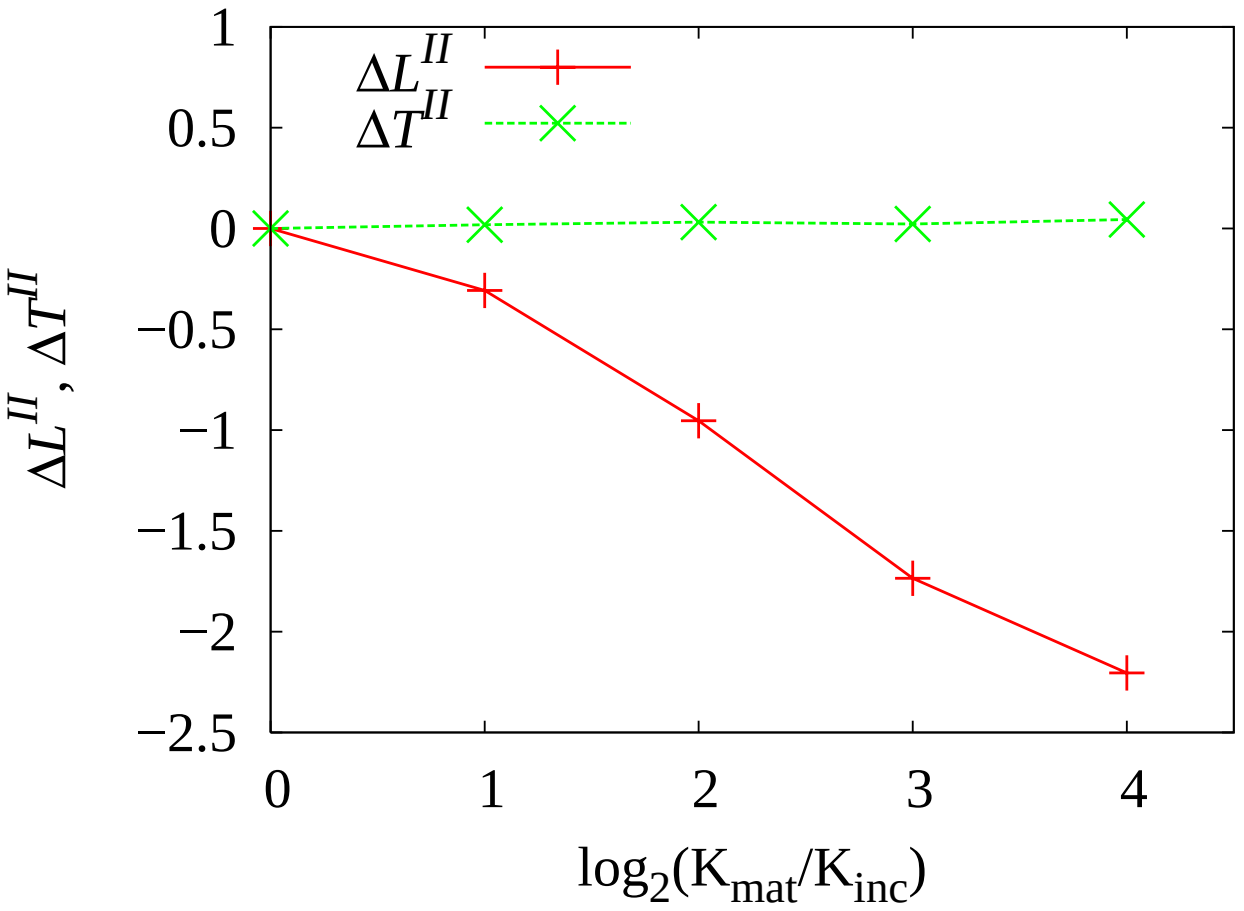

FIG. 6: (color online) Differences $\Delta L^{II}$ (full line with + symbols) and $\Delta T^{II}$ (dashed line with × symbols) between atomistic and continuum results for the longitudinal and transverse nonlinear coefficients, respectively, versus the elastic contrast.

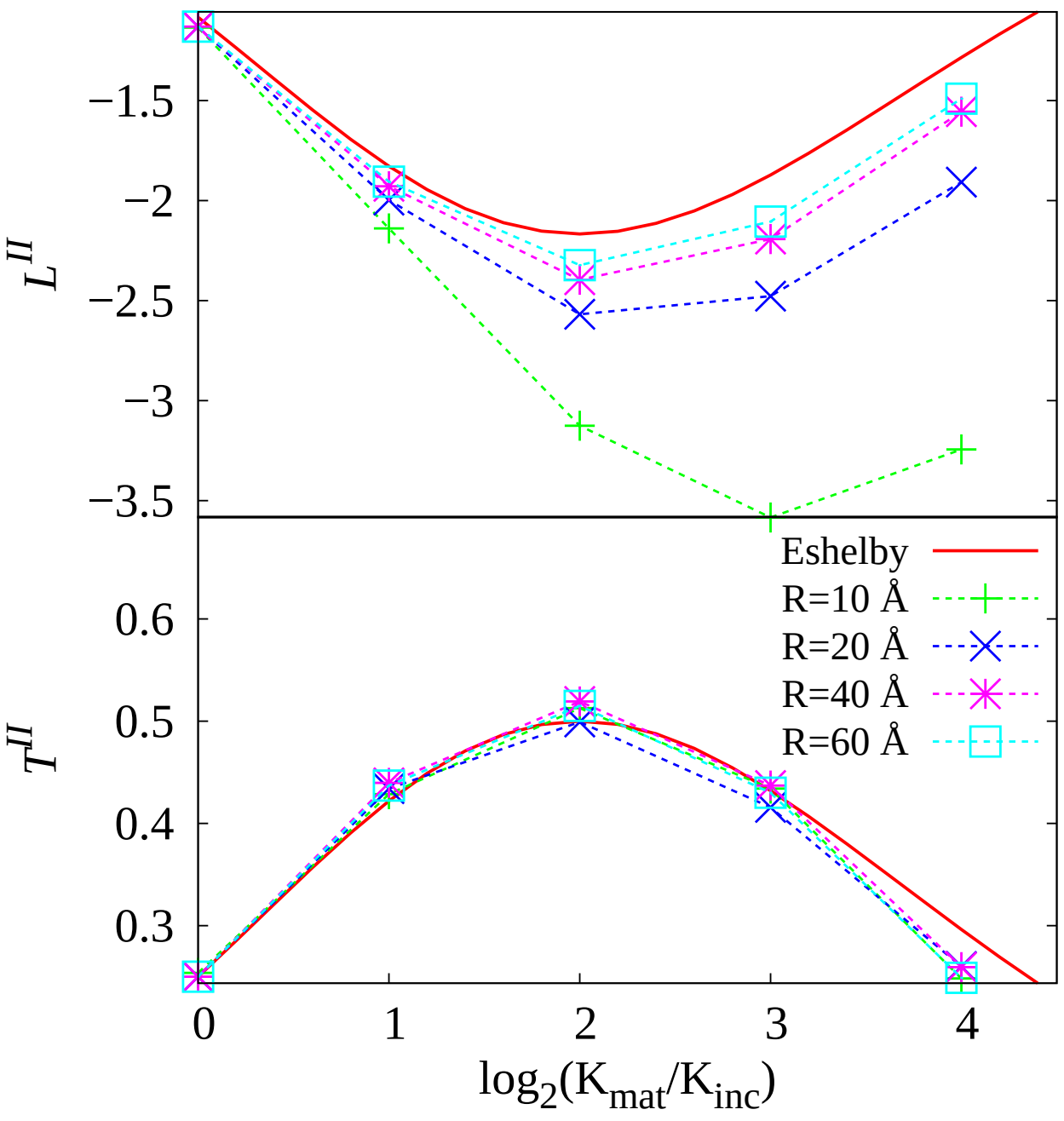

FIG. 7: (color online) Longitudinal (top panel) and transverse (bottom panel) nonlinear coefficient as a function of the elastic contrast, calculated for increasing values of the inhomogeneity radius $R$. The solid lines represent the Eshelby predictions, while dashed lines with symbols represent the atomistic results.

We observe that the atomistic transverse nonlinear coefficient is equal to the corresponding continuum one for any value of the contrast. At variance, sizable discrepancies have been found for the longitudinal coefficient. As shown in Fig.7, these effects depend on the contrast and disappear when the radius of the inhomogeneity increases. Both for the linear and nonlinear coefficients, we observe that the scale effects disappear when the inhomogeneity radius is larger than 10 nm.

### C. Scaling laws for the atomistic effects

We have shown through the previous atomistic simulations that, for a positive elastic contrast between the matrix and the inhomogeneity, the Eshelby theory (both in the linear and nonlinear regime) is recovered only in the limit of a large inhomogeneity. In the present Section, we investigate the scaling laws that drive this phenomenon.

In Fig.8 we report the atomistic results for the longitudinal linear $L^I$ and nonlinear $L^{II}$ coefficients as a function of $R$, normalized by their continuum counterparts $L^I(\infty)$ and $L^{II}(\infty)$. We do not take into consideration the transverse coefficients $T^I$ and $T^{II}$ since they show quite negligible scale effects. Atomistic data are nicely

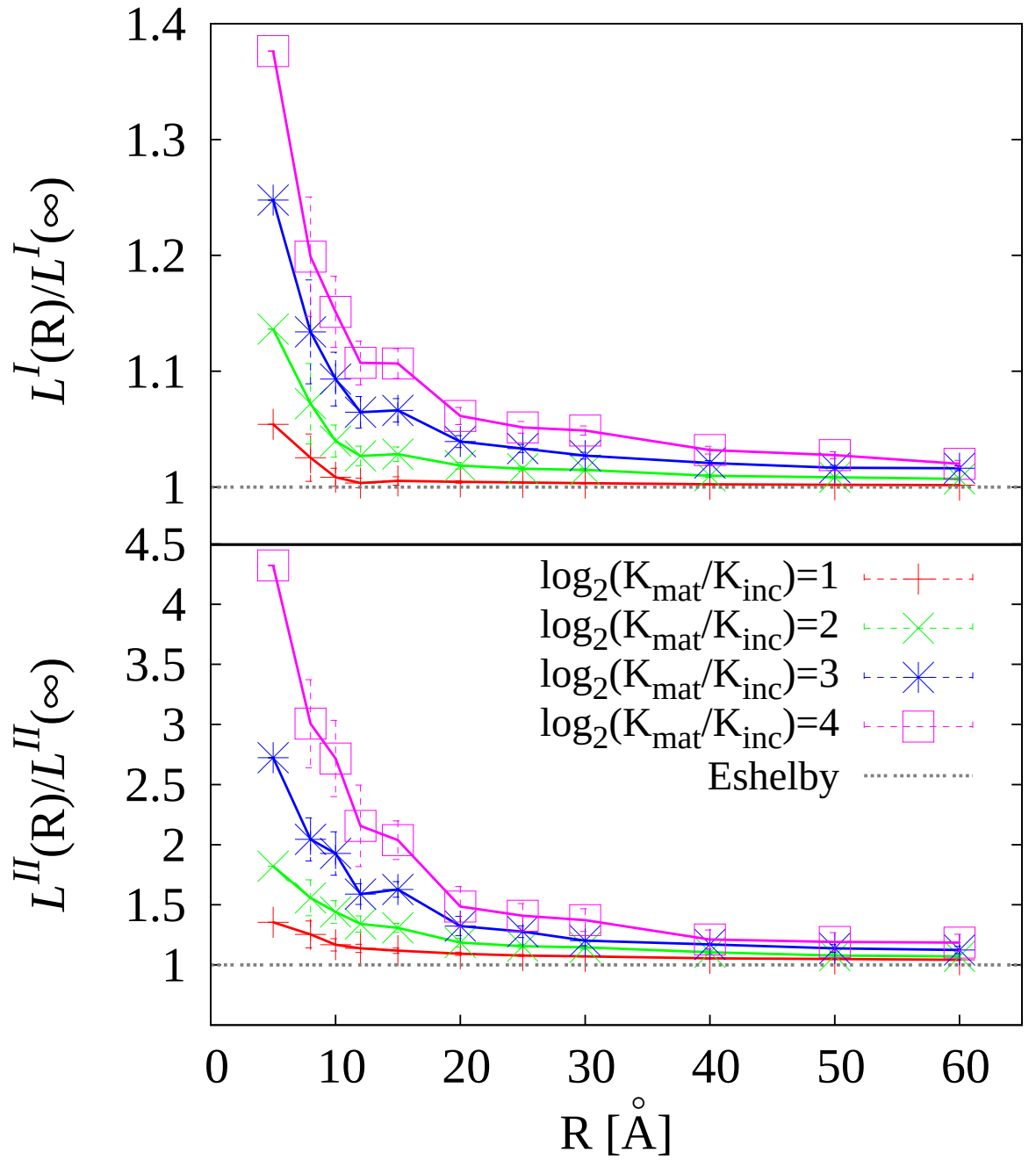

FIG. 8: (color online) Linear $L^I(R)$ (top) and nonlinear $L^{II}(R)$ (bottom) coefficients as function of the radius $R$ of the inhomogeneity, for different elastic contrasts. Both coefficients are normalized to the corresponding values $L^I(\infty)$ and $L^{II}(\infty)$, respectively, predicted by the Eshelby theory.

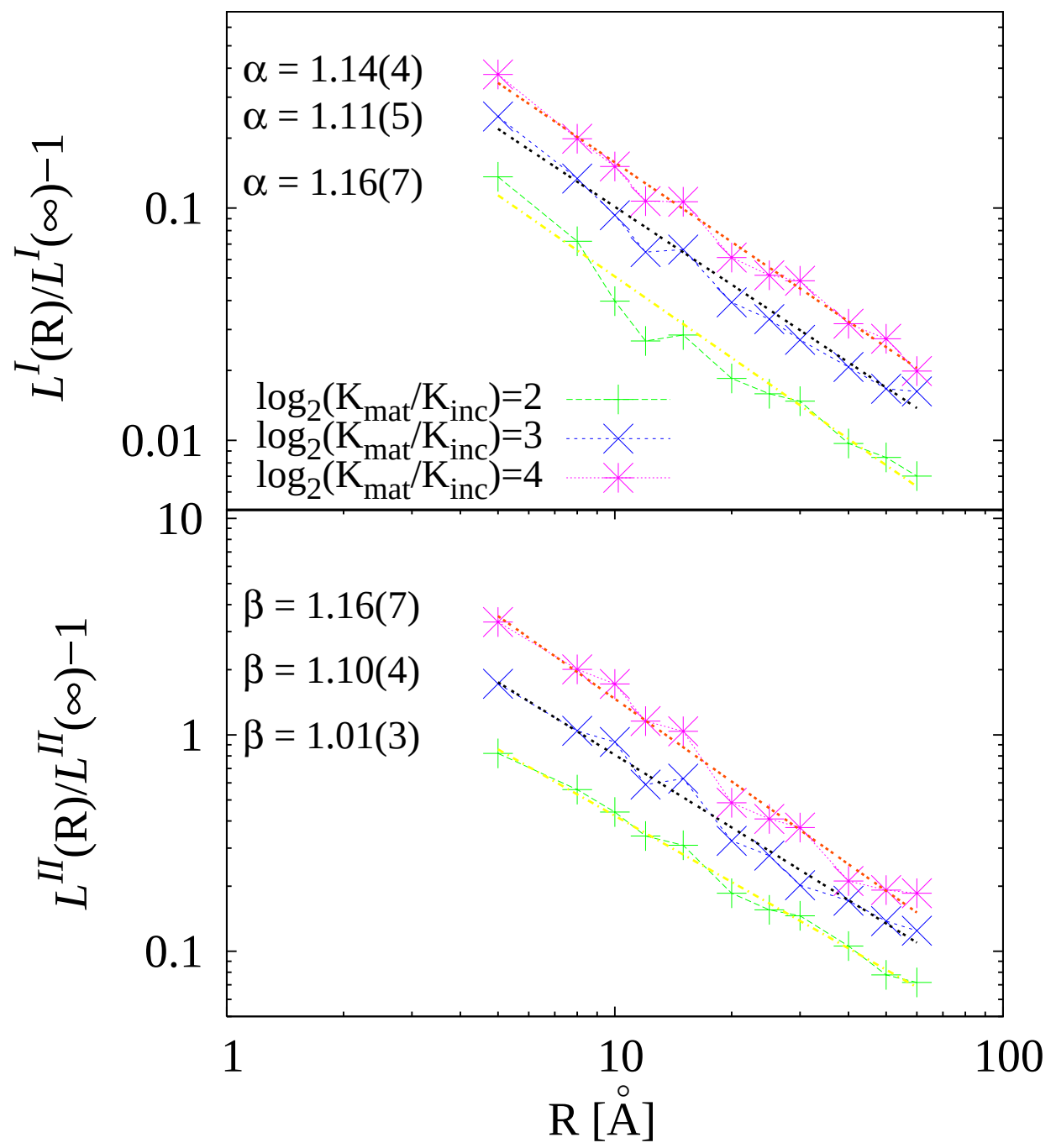

FIG. 9: (color online) Fitting of Eqs.(26) (top) and (27) (bottom) for different elastic contrast between the matrix and the inhomogeneity. The indicated uncertainty (in round brackets) applies to the least significant figure (digit) of the value. For instance, 1.16(7) stands for 1.16 ± 0.07.

fitted by a simple power law as

$$\frac{L^I(R)}{L^I(\infty)} = 1 + \frac{a}{R^\alpha} \qquad (26)$$

$$\frac{L^{II}(R)}{L^{II}(\infty)} = 1 + \frac{b}{R^\beta} \qquad (27)$$

where $a, b, \alpha$ and $\beta$ are fitting parameters. Fig.9 (top) and Fig.9 (bottom) show the result of the numerical fits of Eqs.(26) and (27), respectively, providing the same scaling exponent $\alpha \simeq \beta \simeq 1.1$ for the linear and nonlinear coefficients (with the uncertainties indicated in Fig.9). It is important to remark that the values of these scaling exponents are independent on the elastic contrast, as shown in Fig.9 (top) and Fig.9 (bottom). The above result suggests that the linear and nonlinear behaviors of our lattice system belong to the same universality class. In such a case with $\alpha = \beta$, the overall internal displacement $\varepsilon_l$ fulfills a similar simple power law

$$\varepsilon_l(\epsilon; R) = \varepsilon_l(\epsilon; \infty) + \Delta(\epsilon) R^{-\alpha}$$

where $\Delta(\epsilon) = aL^I(\infty)\epsilon + bL^{II}(\infty)\epsilon^2$. As a consequence of such a scaling behavior, the measurement or the computation of the $\varepsilon_l(\epsilon; \tilde{R})$ curve for a given value of $\tilde{R}$, allows for the direct knowledge of the same curve for an arbitrary radius $R$, being the latter simply proportional to the former

$$\frac{\varepsilon_l(\epsilon; R) - \varepsilon_l(\epsilon; \infty)}{\varepsilon_l(\epsilon; \tilde{R}) - \varepsilon_l(\epsilon; \infty)} = \left(\frac{R}{\tilde{R}}\right)^{-\alpha}$$

In other words, because of the relation $\beta = \alpha$, nano-inhomogeneities with different radii exhibit responses to the external load which differ only for a constant scale factor $(R/\tilde{R})^{-\alpha}$ independently on the magnitude $\epsilon$ of the applied strain. These conclusions have been proved within the range 5 Å - 60 Å for the radius oh the inhomogeneity.

The results here obtained can be compared with those discussed in a recent analysis concerning the scaling law for properties of nano-structured materials.[36] In this work the elasticity of a non-ideal surface is characterized by two surface elastic constants giving rise to two intrinsic length scales. Thus, the size-dependence of physical properties associated with the deformation problems of heterogeneous nano-solids is expected to follow a scaling law with an intrinsic length scale which is a linear combination of these two scales. This approach is based on the so-called Interface Stress Model (see Introduction) and it leads to the scaling exponent $\alpha = 1$ for the linear properties. This value can be explained through the competition between the elastic surface energy at the interface and the strain energy in the bulk. Our value $\sim$1.1 of the scaling exponent suggests that the onset of the length scale dependence in the elastic behavior of the inhomogeneity cannot be explained just in terms of the sole competition between surface and volume energies. Therefore, we suggest that such a scale dependence should be rather related to the discretization of the continuum equations at the atomic scale introduced by the present lattice model. In fact, in our model, the lattice is perfect in the whole plane (with no structural interface mismatch) and the interface concept is introduced only by setting different linear and nonlinear spring constants in the internal (inhomogeneity) and external (matrix) regions. In order to better explain the differences of the results between the application of the Interface Stress Model[36] and our lattice model, we also remark that the first one is expected to give good predictions for a typical size of the particle larger than 3-5 nm, while the present approach, as above said, has been applied in the range 5 Å - 60 Å. It means that scale effects induced by disordered interfaces or interface mismatches are typically exhibited at a length scale larger than the scale effects induced by the lattice structure of the matter (taken into account simply by its interatomic distance).

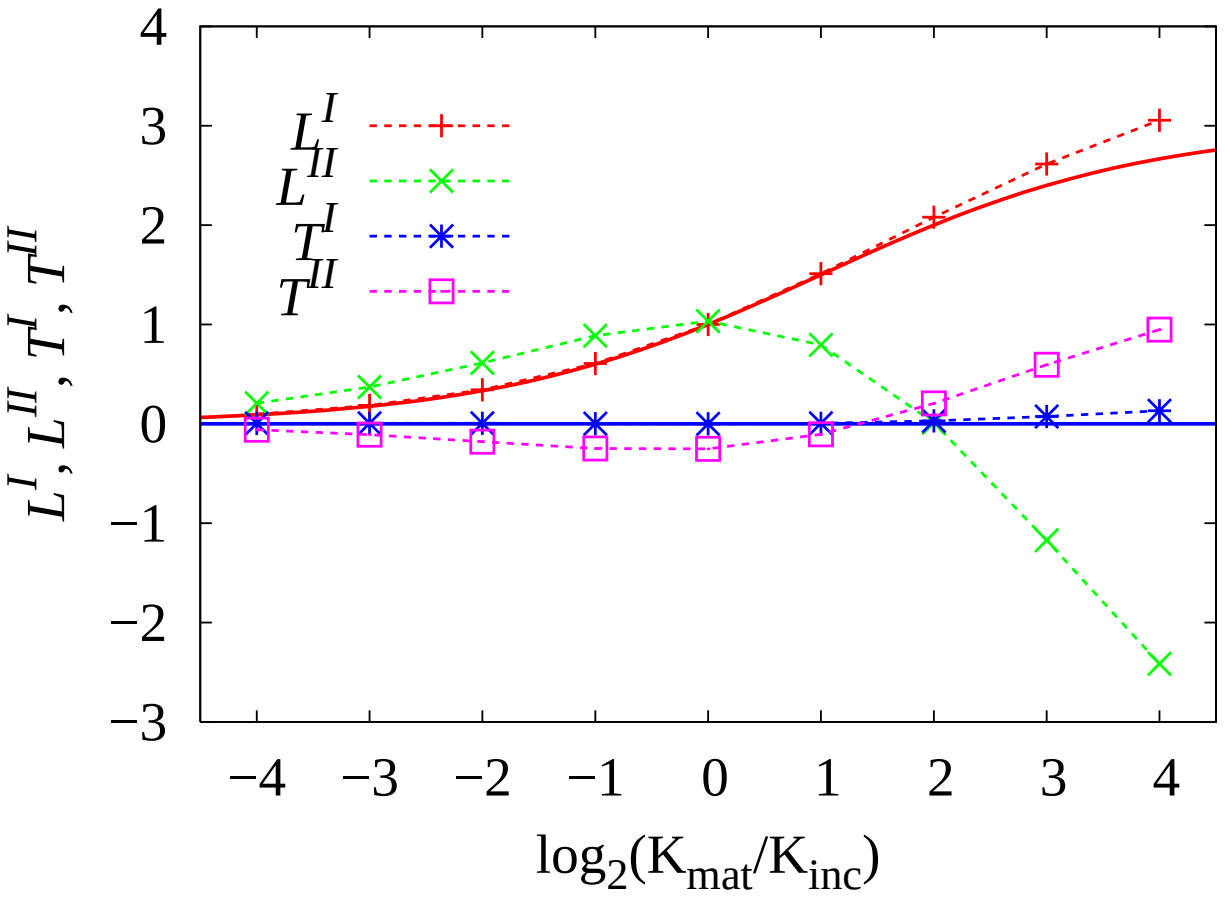

FIG. 10: (color online) Elastic behavior of a linear inhomogeneity embedded in a nonlinear matrix under remote load as summarized by the linear ($L^I$ and $T^I$) and nonlinear ($L^{II}$ and $T^{II}$) coefficients. The solid lines represent the linear results of the continuum Eshelby theory for $L^I$ and $T^I$ given in Eqs.(20) and (21). Dashed lines with $+$ and $\times$ ($\Box$ and $*$) symbols represent the atomic result for the longitudinal (transverse) coefficients.

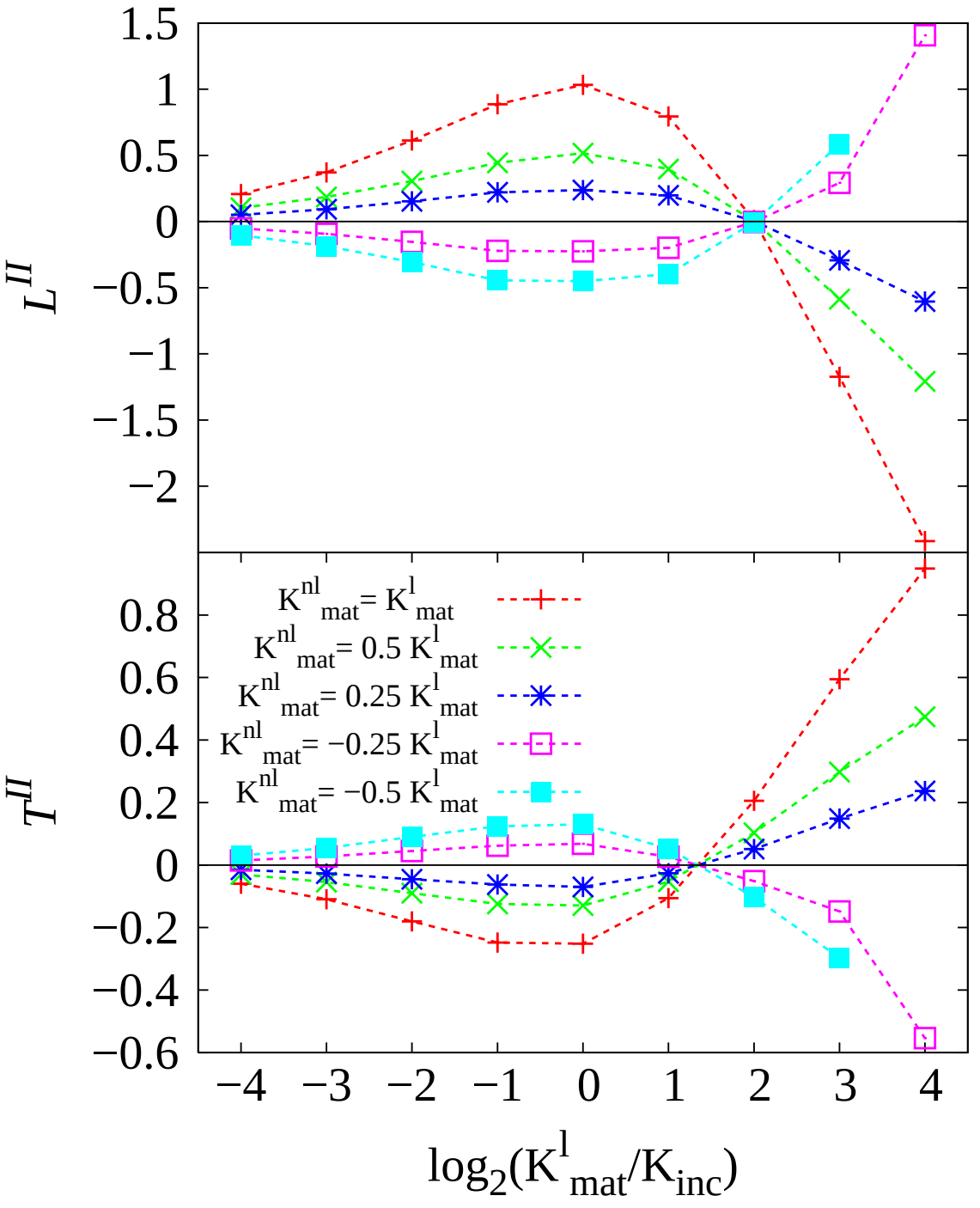

FIG. 11: (color online) Atomistic results for the nonlinear coefficients $L^{II}$ (top) and $T^{II}$ (bottom) versus the elastic contrast $\log_2(K^l_{mat}/K_{inc})$ for different values of the nonlinearity ratio $K^{nl}_{mat}/K^l_{mat}$ in the matrix.

### D. Linear inhomogeneity in a nonlinear matrix

In this Section we consider the case of a linear inhomogeneity embedded into a nonlinear matrix. The elastic behavior of the inhomogeneity is described by the Eq.(24) with $K = K_{inc}$, while the matrix is modeled according to Eq.(25) with $K = K_{mat}$.

In Fig.10 we report the longitudinal and transverse coefficients for several values of the elastic contrast. Even in this case the internal strain field is found to be uniform, although this result is not anticipated by the Eshelby theory. By comparing Fig.10 and Fig.5, we can state that when the inhomogeneity is nonlinear then the coefficients $L^{II}$ and $T^{II}$ have constant sign, independently on the contrast. Furthermore, they exhibit a minimum and a maximum, respectively (see Fig.5). On the contrary, when the matrix is nonlinear it is remarkable to observe that two values of contrast exist which cancel out the second order nonlinear effects in the longitudinal and transverse direction, respectively (Fig.10).

As mentioned above, to the best of our knowledge a continuum theory solution of this case is not available. Therefore, we further analyze the elastic behavior of the inhomogeneity/matrix system by varying the nonlinearity of the matrix. To this aim, we have set $\kappa_l = K$, $\kappa_h = K'$ and $\kappa_a = \frac{3}{2}K'$ within the nonlinear matrix, where $K$ and $K'$ are constants, so to get

$$\begin{aligned} C^{nl}_{11} &= \frac{3\sqrt{3}}{4}K^{l}_{mat} \\ C^{nl}_{12} &= \frac{\sqrt{3}}{4}K^{l}_{mat} \\ C^{nl}_{111} &= \frac{9}{4}\sqrt{3}K^{nl}_{mat} \\ C^{nl}_{222} &= \frac{9}{4}\sqrt{3}K^{nl}_{mat} \\ C^{nl}_{112} &= \frac{\sqrt{3}}{4}K^{nl}_{mat} \end{aligned} \qquad (28)$$

where $K^{l}_{mat} = K + K'$ and $K^{nl}_{mat} = K'$ directly affect the linear and nonlinear behavior, respectively. By varying the value of $K^{nl}_{mat}$ with respect to $K^{l}_{mat}$, we can emphasize the nonlinear regime. In Fig.11 we report the atomistic results for the nonlinear coefficients $L^{II}$ (top) and $T^{II}$ (bottom) versus the (linear) elastic contrast for different values of nonlinearity ratio $K^{nl}_{mat}/K^{l}_{mat}$ in the matrix. We have not reported the results for the linear coefficients $L^{I}$ and $T^{I}$ since they are not affected by the nonlinear features of both inhomogeneity and matrix; indeed, they assume the very same values reported in Fig.10. It is interesting to underline that the longitudinal coefficient $L^{II}$ vanishes for a given linear contrast for any possible value of the nonlinear parameter $K^{nl}_{mat}$ of the matrix. The same phenomenon has been observed for the transverse coefficient $T^{II}$.

### E. Nonlinear inhomogeneity in a nonlinear matrix

In this Section we finally consider the case of a nonlinear inhomogeneity embedded into a nonlinear matrix. We start by considering both media as described in Eq.(25) with $K = K_{inc}$ and $K = K_{mat}$, respectively in the inhomogeneity and matrix. In Fig.12 we report the longitudinal and transverse coefficients for several values of the elastic contrast. In this case the zero contrast value corresponds to a nonlinear but homogeneous material (without inhomogeneity). Therefore, we obtained $L^{II} = T^{II} = 0$ and $L^{I} = 1$ for $K_{mat} = K_{inc}$, as expected.

Since, as in the previous case, this configuration is hardly affordable by continuum theory, we performed a more detailed analysis by investigating several nonlinear hosting matrices, all modeled by Eq.(28) but characterized by a different ratio $K^{nl}_{mat}/K^{l}_{mat}$ between the nonlinear and the linear stiffness coefficients. We have instead set the behavior of the inhomogeneity according to Eq.(25), with $K = K_{inc}$.

In Fig.13 we show the atomistic results for the nonlinear coefficients $L^{II}$ (top) and $T^{II}$ (bottom) versus the elastic contrast for different values of the $K^{nl}_{mat}/K^{l}_{mat}$ ratio. We have not reported the results for the linear coefficients $L^{I}$ and $T^{I}$ since they are not affected by the nonlinear features of both inhomogeneity and matrix; indeed, they assume the very same values reported in Fig.12. Interesting enough, we observe that there is a value of the (linear) elastic contrast $\log_2(K^{l}_{mat}/K_{inc})$ which generates a constant value of $L^{II}$ (see Fig.13, top) for any nonlinearity of the matrix. This result indicates that, in such a specific condition, the nonlinear effects of the matrix are quenched. The same behavior is also observed for the transverse coefficient $T^{II}$ (see Fig.13, bottom).

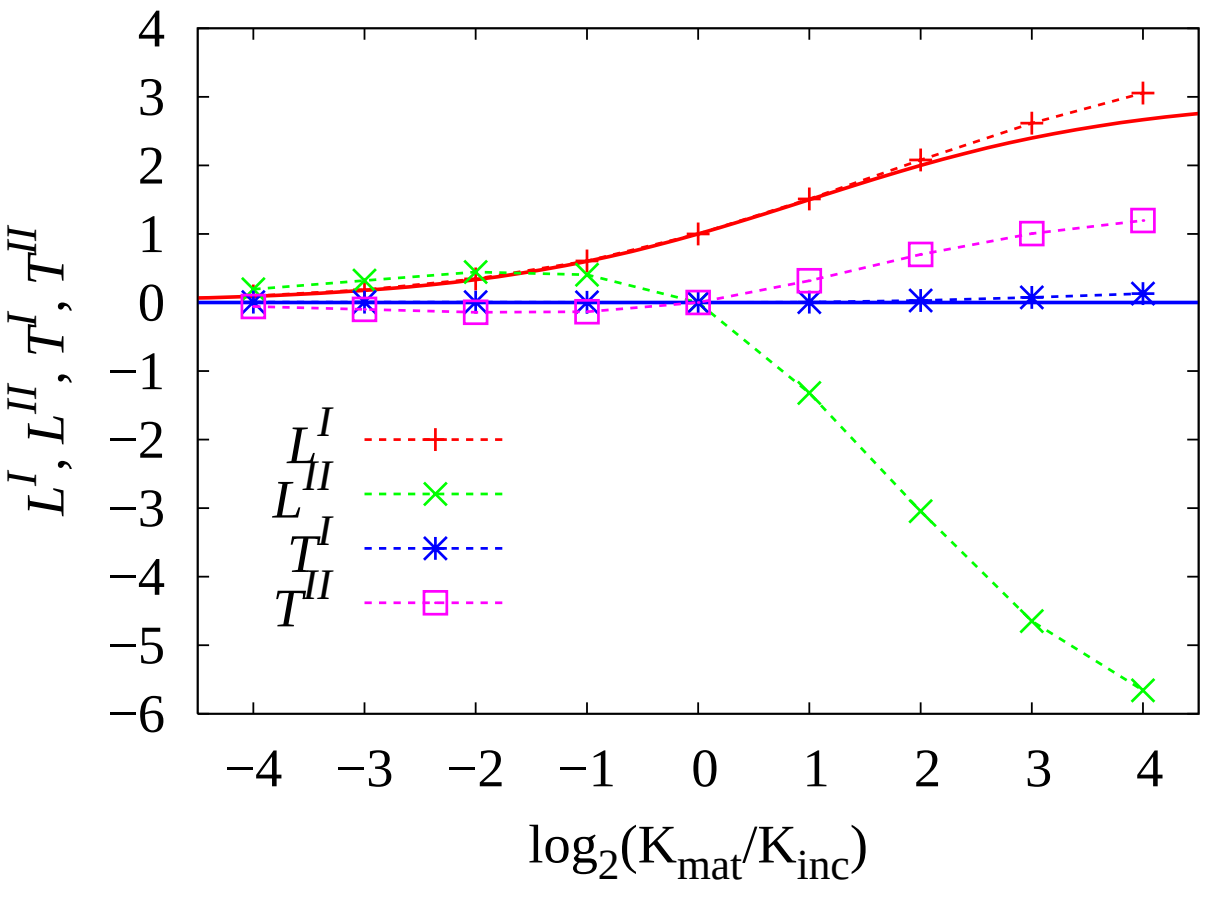

FIG. 12: (color online) Elastic behavior of a nonlinear inhomogeneity embedded in a nonlinear matrix under remote load as summarized by the linear ($L^{I}$ and $T^{I}$) and nonlinear ($L^{II}$ and $T^{II}$) coefficients. The solid lines represent the linear results of the continuum Eshelby theory given in Eqs.(20)-(21). Dashed lines with + and × (□ and ∗) symbols represent the atomic result for the longitudinal (transverse) coefficients.

## V. CONCLUSIONS

In this work we have introduced a conceptual mapping of the constitutive linear and nonlinear equations of the continuum elasticity theory onto a lattice model exploiting the real atomistic structure of an embedded nano-inhomogeneity. The present lattice model naturally introduces the notion of length-scale and, therefore, opens the possibility to investigate by computer experiments possible scale effects on the elastic behavior of nanostructured materials. We have thoroughly applied the elastic model to investigate nonlinear inhomogeneities and matrices, also addressing some configurations which can hardly treated by continuum theory.

Firstly, we have proved that the atomistic lattice model is in perfect agreement with the Eshelby theory for linear or nonlinear large inhomogeneities embedded in a linear matrix. When the radius of the inhomogeneity becomes comparable with the interatomic distance of the involved materials, the scale-effects drive the elastic features, exhibiting sizable deviations from the continuum results. More specifically, we have observed that such effects are stronger for a positive elastic contrast, i.e. for a matrix

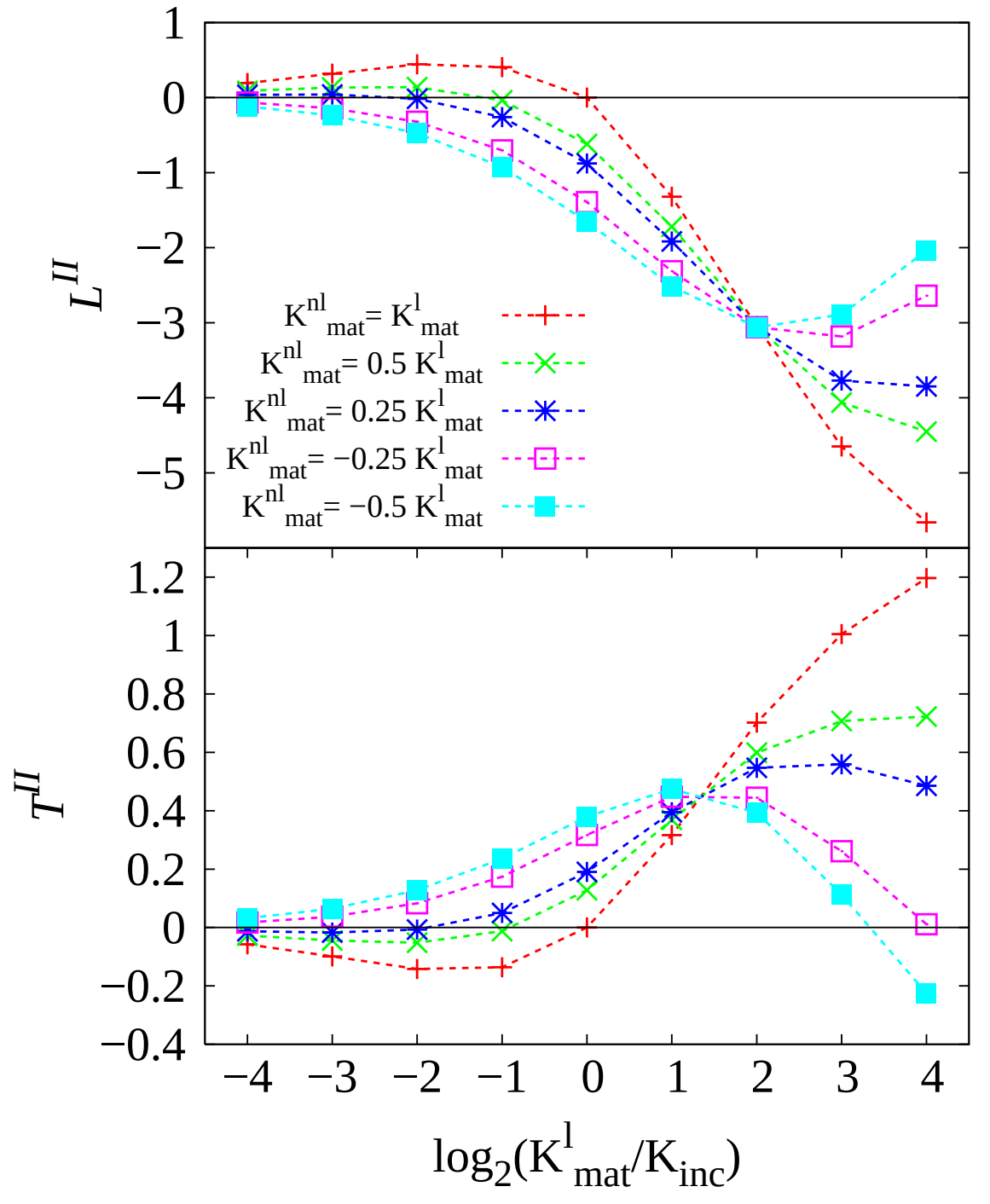

FIG. 13: (color online) Atomistic results for the nonlinear coefficients $L^{II}$ (top) and $T^{II}$ (bottom) versus the (linear) elastic contrast $\log_2(K^l_{mat}/K_{inc})$ for different values of the $K^{nl}_{mat}/K^l_{mat}$ ratio in the matrix.

stiffer than the inhomogeneity.

Secondly, we have investigated the case of a nonlinear matrix, embedding either a linear or a nonlinear inhomogeneity, by calculating the corresponding elastic fields. In particular, in the case of a linear inhomogeneity we have proved that the nonlinear response vanishes for a given linear contrast (for any nonlinearity of the matrix).

Finally, we have proved that linear and nonlinear scale effects are described by the same scaling exponent, independently of the elastic contrast. This suggests that the overall strain (or stress) field within the inhomogeneity can be described by similar power law with the same scaling exponent.

In conclusion, the present lattice model -which is computationally not intensive and very easy to implement- is proposed as a valuable theoretical tool for investigating the combination of scale-effects and nonlinear elasticity in arbitrarily complicated nanostructured materials (such as e.g.: nano-alloys, nano-composites and nano-graded interfaces).

### Acknowledgments

This work is founded by "Regione Autonoma della Sardegna" under project "Modellazione Multiscala della Meccanica dei Materiali Compositi (M4C)". One of us (S. G.) acknowledges financial support by CYBERSAR (Cagliari, Italy). We also acknowledge computational support by CYBERSAR and by CASPUR (Rome, Italy) under project "Standard HPC grant 2009".

---

[*] Electronic address: `pierluca.palla@dsf.unica.it`
[†] Electronic address: `stefano.giordano@dsf.unica.it`
[‡] Electronic address: `luciano.colombo@dsf.unica.it`